\documentclass[aps,prl,twocolumn,floats,showpacs,floatfix,a4paper]{revtex4}
\usepackage{epsfig}
\usepackage{color}
\usepackage{bm}
\usepackage{latexsym}

\begin{document}

\bibliographystyle{unsrt}

\newcommand{\half}{\mbox{\small $\frac{1}{2}$}}
\newcommand{\bra}[1]{\left\langle #1\right|}
\newcommand{\ket}[1]{\left|#1\right\rangle }
\newcommand{\acos}[1]{\mbox{a}\cos }

%%%%%%%%%%%%%%%%%%%%%%%%%%%%%%%%%%%%%%%%%%%%%%%%%%%%%%%%%%%%%%%%%%%%%%%%
\title{Scaling properties of delay times in one-dimensional random media}
\author{Joshua D. Bodyfelt$^{1}$, J. A. M\'endez-Berm\'udez$^{2}$, Andrey Chabanov$^{3}$, Tsampikos Kottos$^{1}$}

\affiliation{$^1$Department of Physics, Wesleyan University, Middletown, Connecticut 06459, USA \\
$^2$Instituto de F\'{\i}sica, Universidad Aut\'onoma de Puebla, Apartado Postal J-48, Puebla 72570, Mexico\\
$^3$Department of Physics and Astronomy, The University of Texas at San Antonio, Texas 78249 USA}

\begin{abstract}
The scaling properties of the inverse moments of Wigner delay times are investigated in finite 
one-dimensional ($1D$) random media with one channel attached to the boundary of the sample. 
We find that they follow a simple scaling law which is independent of the microscopic details of 
the random potential. Our theoretical considerations are confirmed numerically for systems as diverse as
$1D$ disordered wires and optical lattices to microwave waveguides with correlated scatterers.
\end{abstract}


\maketitle

%%%%%%%%%%%%%%%%%%%%%%%%%%%%     Introduction     %%%%%%%%%%%%%%%%%%%%%%%%%%%%%%%%%%%%%
%\section{Introduction}

The study of the statistical properties of Wigner delay times has been a subject of intense research 
activity \cite{FS97,vBSB01,K05,OF05,TC99,F03,MBK05,KW02,OKG00,GSST99,CG01,PJM03,SvBB00,MFME06}. The 
Wigner delay time is defined as the energy derivative of the total phase of the scattering matrix $S$ 
\textit{i.e.}, $\tau_W=-i\hbar \partial \ln \det S/\partial E$, and can be interpreted as a time delay 
in propagation of the peak of the wave packet due to scattering interference, in comparison to a free 
wave packet propagation. Although most of the contemporary activity has been focused in understanding 
the statistical properties of delay times within chaotic mesoscopic systems \cite{FS97, vBSB01}, recently 
the interest has shifted towards random scattering media exhibiting Anderson localization \cite{K05,
OF05,TC99,F03,OKG00,GSST99,CG01,PJM03, SvBB00} including the most difficult case of the Anderson 
Metal-Insulator Transition (MIT) \cite{K05,KW02,MBK05,OF05,MFME06}. On the experimental side \cite{GSST99,
CG01,PJM03}, the statistics of scattering phases and delay times have been measured in microwave experiments 
with quasi-one-dimensional random samples, while on the theoretical side the main effort has been to connect 
the statistical properties of delay times with that of eigenfunctions
\cite{K05,MBK05,OF05,MFME06}. Establishing such a relation may open new exciting opportunities 
for measuring the statistical properties of eigenfunctions \cite{CGIFM92,FM92,FM94,AKL91,MK95,
FE95,M97,SA97,ARS02,OKG02} via the experimentally accessible delay times.

Specifically, using the powerful Non-Linear $\sigma$ Model (NL$\sigma$M) technique \cite{OF05,MFME06} 
an exact relation was found linking the probability distribution of eigenfunction 
components within a random medium to the distribution of Wigner delay times in the same sample of 
length $L$, with one channel attached at its bulk. This relation is {\it exact on the level of 
the NL$\sigma$M} and valid independent of the system size $L$ (\textit{i.e.} irrespective if we take the 
thermodynamic limit $L\rightarrow \infty$ or keep $L$ finite). 

However, one has to question the validity of mapping a particular microscopic model of a 
disordered system onto the NL$\sigma$M. More specifically, this mapping is approximately correct 
in the case of weak disorder and breaks down totally for strong disorder. Another strict 
requirement is that the underlying geometry allows for a diffusive process - this certainly 
is not the case for strictly one-dimensional ($1D$) random media. Finally, NL$\sigma$M calculations 
pre-assume that the disorder potential is white-noise, thus excluding the emerging family of disordered 
systems with imprinted correlations in their potential \cite{IK99,KIKS00,DWP90,ML98,RMSDLA04,
BDHTTPAA99, KIK07, TS05}. The above restrictions cast reasonable doubts on the validity of NL$\sigma$M
predictions, as far as {\it realistic} systems are concerned, and call for testing by means of 
a dedicated experiment or computer simulation.

It is the purpose of this paper to investigate the scaling properties of moments of delay 
times and compare them with the ones found for wavefunctions in cases where the conditions 
for NL$\sigma$M applicability are violated. To this end we will study various microscopic 
systems: (a) a $1D$ disordered electronic system (modeled by an Anderson Hamiltonian), (b) 
a microwave system with long-range correlated 
scatterers inside a waveguide (modeled by a Kronig-Penney model), and (c) cold-atoms in a 
disordered optical lattice (modeled again by a Kronig-Penney model with binary distribution). 
In all cases, we find that the inverse moments of delay times $\tau_L$ in a disordered sample
of length $L$ follow a simple scaling law which is independent of the microscopic properties 
belonging to the underlying physical system. Specifically we find that
\begin{equation}
\beta_{-q}=f(\lambda_{-q}) \quad ; \quad \beta_{-q} \equiv \frac{\tau_{\rm ref}^{-q}} 
{\langle \tau_L^{-q} \rangle },\quad \lambda_{-q} \equiv \frac{\tau_{\rm 
ref}^{-q}} {\langle \tau_{\infty}^{-q} \rangle },
\label{tscale1}
\end{equation}
where $q$ takes positive values and $\langle \tau_L\rangle$ represents the average (or typical) 
delay time over disorder realizations. The variable $\tau_{\infty}$ 
represents the delay time of the $L\rightarrow \infty$ sample with the same disordered potential. 
The variable $\tau_{\rm ref}$ is the delay time of a "reference" sample, corresponding to an "infinite" 
localization length set-up, with length $L$. The former quantity incorporates the microscopic 
information of the system (\textit{i.e.} disorder potential), whereas $\tau_{\rm ref}$ only depends
on the information of the finite sample length $L$, as well as the dimensionality and the energy $E$ 
at which the scattering experiment is performed. Our numerical analysis indicates that the scaling 
law, Eq.(\ref{tscale1}), can take the model-independent form
\begin{equation}
\langle \tau^{-q}_L(\epsilon, E) \rangle = \langle \tau^{-q}_\infty(\epsilon, E) 
\rangle + \tau_{\rm ref}^{-q}
\label{tscale3}
\end{equation}
where $\epsilon$ is the disorder strength of the random potential. In fact, our numerical data, 
suggest that Eq.(\ref{tscale3}) is exact only for $q=1$ while for higher $q-$values small deviations
from the linear behaviour can be detected. 

We point out that a similar relation to Eq.(\ref{tscale3}) was found for the scaling properties 
of wavefunction moments within a closed disordered sample \cite{CGIFM92,FM92,FM94}. The corresponding 
expression involves the $q'=q+1$ wavefunction moment and reads
\begin{equation}
{1\over \langle l^{(q')}_L(\epsilon, E) \rangle} = {1\over \langle l_\infty^{(q')}(\epsilon, E) \rangle} 
+ {1\over l^{(q')}_{\rm ref}}
\label{yscale1}
\end{equation}
where $l^{(q')}_L=L (P^{(q')}/P^{(q')}_{\rm ref})^{1/(1-q')}$ are the various information lengths 
of a sample with length $L$, $P^{(q')}\equiv \sum_n^L |\psi_n|^{2q'}$ with eigenfunction components 
$\psi_n$, $l_\infty(\epsilon, E)$ is the localization length of the infinite sample with the same 
disordered strength, and $P^{q'}_{\rm ref}\sim L$ with a pre-factor defined by the reference geometry 
($1D$ periodic lattice in the cases studied here). For the special case $q'=1$ the corresponding information 
length is equal to the entropic length defined by $l^{(1)}_L = e/2\exp(-\sum_{n=1}^L |\psi_n|^2 
\ln|\psi_n|^2)$.

We initiate our analysis by recalling the notion of delay times as originally proposed by Wigner
\cite{W55,CN02}. This is the time that a reflected particle is delayed due to interaction with
a scattering region. Now we recall that the $q'=2$ information length, $l_L^{(2)}(\epsilon, E)$,
(associated with the inverse participation ratio) measures the "penetration" (localization) 
length inside a disordered sample before the particle is reflected back (we are considering 
here the one channel scattering set-up). The corresponding delay time due to the scattering 
from the disordered sample is then given by $\tau_L = 2l^{(2)}_L /v$, where $v$ is the group 
velocity of the wavepacket centered around energy $E$. Using this argument and substituting 
it for $l^{(2)}_L$ in Eq.(\ref{yscale1}), we obtain Eq.(\ref{tscale3}) for $q=1$. In fact, our 
numerical data (see below) indicate that Eq. (\ref{tscale3}) describes to a good approximation 
higher $q$-moments as well.

Below we report our numerical results for various microscopic models which support the scaling 
of Eq.(\ref{tscale1}). Although our presentation focuses on the first moment $q=1$, we have 
found that higher moments follow the scaling law, Eq.(\ref{tscale1}), equally well.

%%%%%%%%%%%%%%%%%%%%%%%%%%%%%%%%%%%%%%%%%%%%%%%%%%%%%%%%5
{\it $1D$ Disordered Electronic System -} 
The standard model that describes a one-dimensional 
disordered electronic sample is the tight-binding equation
\begin{equation}
\psi_{n+1}+\psi_{n-1} = \left( E(k)-V_n \right) \psi_n;\quad n=1,2,\cdots,L
\label{TBE}
\end{equation}
where $k$ is the incident wavenumber and $\psi_n$ is the wavefunction amplitude at the 
$n^{\rm th}$ site. The on-site potential $V_n$ for $1\leq n\leq L$ is independently and 
identically distributed with a box probability distribution, \textit{i.e.} the $V_n$ are 
uniformly distributed on the interval $[-\epsilon/2, \epsilon/2]$. 

\begin{figure}[htb]
\includegraphics[width=\columnwidth,keepaspectratio,clip]{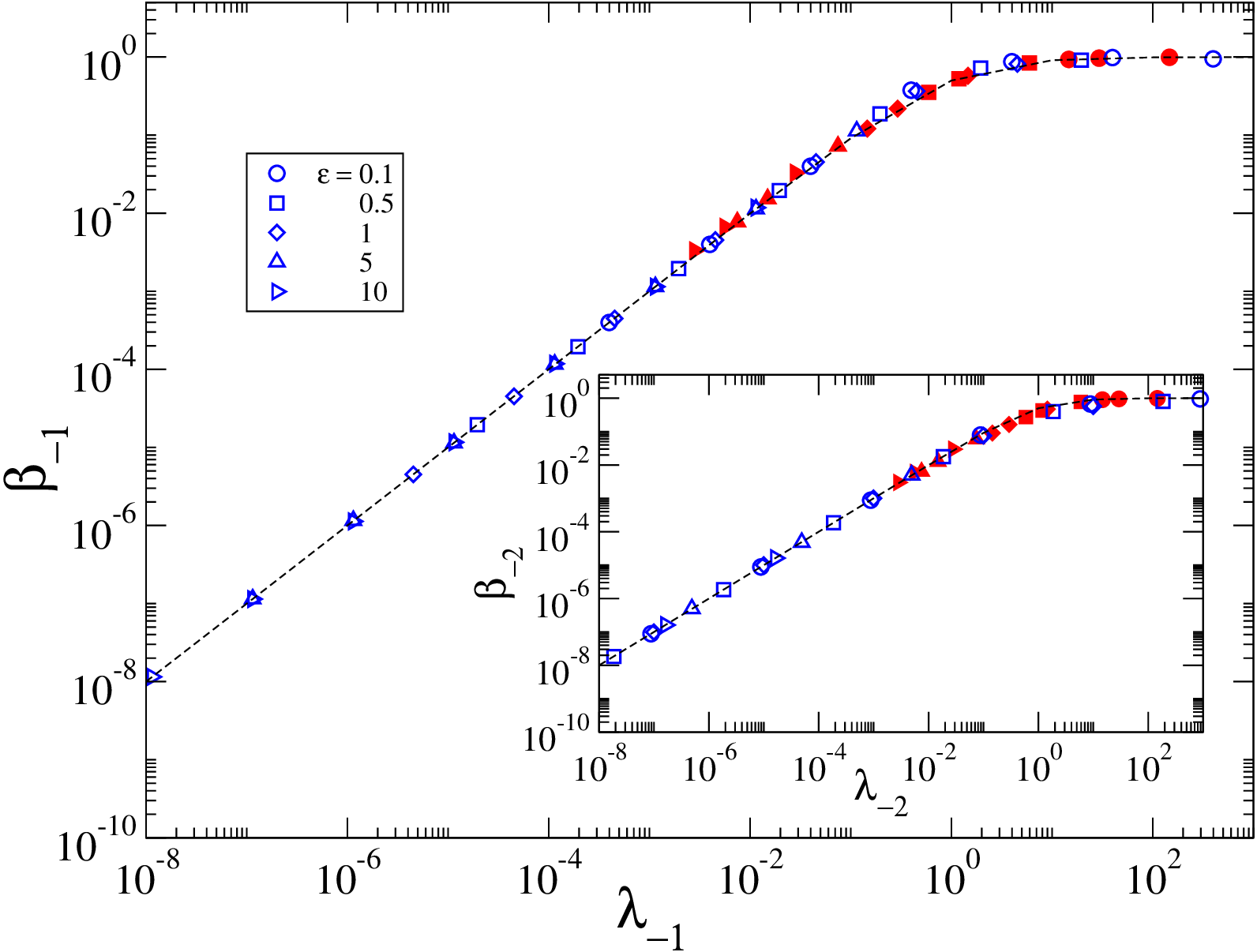}
\caption{(Color online) Scaled inverse delay times,Eq.(\ref{tscale1}), for the
Anderson model. Various symbols correspond to different disordered potentials 
$\epsilon \in \left\lbrace 0.1,0.5,1,5,10 \right\rbrace$ and $|E(k=\sqrt{\pi})|<1$.
Blue hollow symbols denote delay time data for $q=1$.
For comparison, red solid symbols denote $q'=2$ information length data, \textit{i.e.} 
$l_L^{(q')}/ L$ versus $\lambda= 2l_{\infty}(\epsilon,E)/L$.
The dashed line is the result of the best fit from Eqs.(\ref{XYvar},\ref{XYline}).
Inset: Same as in the main figure but now $q=2$ for delay times and $q'=3$ for information lengths.
}\label{AnderScale}
\end{figure}

We open the sample by attaching one channel to the first site $n=1$. The Wigner delay time 
of a sample of length $n+1$ is then evaluated with the use of the Hamiltonian map approach 
\cite{OKG00} through the following iteration relations
\begin{eqnarray}
\tau_{n+1} &= G_n^{-1} \left( \tau_n + \frac{1}{\sin k} \right) + \frac{A_n}{1+
\left[ \tan(\phi_n-k)+A_n\right] ^2} \frac{\cot k}{\sin k}, \label{Taun} \\ 
G_n &= 1 + A_n\sin\left[ 2(\phi_n-k)\right] + A_n^2\cos^2(\phi_n-k), \nonumber
\end{eqnarray}
where $A_n= {V_n\over \sin k}$ and the scattering phase is given by
\begin{eqnarray}
\tan(\phi_{n+1}) = \tan(\phi_n-k)+ A_n. \label{Phin}
\end{eqnarray} 
In Fig.~\ref{AnderScale} we report the delay times for the Anderson model, Eq.(\ref{TBE}). The 
data are averaged over an ensemble of $10^4$ realizations of the random potential and are 
plotted according to the scaling, Eq.(\ref{tscale1}). The value of $\langle \tau_{\infty}^{-q}
\rangle$ was calculated for a sample of length $L = 10^7 $ and its convergence was checked by 
increasing the system size by an additional order i.e. $L=10^8$. The scaled data - for various 
$L$'s and disordered strengths $\epsilon$ - falls on a single curve, confirming the validity of 
the theoretical prediction, Eq.(\ref{tscale1}). Within the same figure we also report the 
corresponding scaled entropic lengths (see solid red symbols)
$l^{(q')}_L/L$ versus the localization parameter $\lambda= 2l_{\infty}(\epsilon,E)/L$, in order
to compare with the scaling law that dictates the delay times. The agreement between information 
lengths and delay times is evident, thus confirming that these two quantities are directly related.
In the inset of Fig.~\ref{AnderScale} we also report our numerical results for the second 
moment, \textit{i.e.} $q=2$. A nice agreement with the theoretical expectation, Eq.(\ref{tscale1}), 
is again quite evident.

%%%%%%%%%%%%%%%%%%%%%%%%%%%%%%%%%%%%%%%%%%%%%%%%%%%%%%%%
{\it Microwaves propagating in a $1D$ waveguide -} 
The creation of frequency pass/stop bands separated 
by mobility edges and their manipulation by imposing appropriate correlations in the disordered
potential \cite{IK99,KIKS00,DWP90,ML98,RMSDLA04,BDHTTPAA99} have recently gained considerable research
interest due to their immediate technological applications. One prominent 
theoretical suggestion \cite{IK99} was based on the introduction of long-range correlations in 
the on-site disordered potential. The theoretical predictions were further supported by
subsequent experimental microwave measurements \cite{KIKS00}, carried out in a single-mode waveguide 
with correlated scatterers realized by screws extending from a waveguide wall. 
By arranging the lengths of the screws according 
to a predefined sequence, correlated scattering arrangements could be realized leading to predefined
mobility edges. If the screws are approximated by delta scatterers, the propagation of a single 
mode in the waveguide can be described by the wave equation for the Kronig-Penney model
\begin{equation}
\psi ''(z)+E\psi (z) = \sum_{n=-\infty}^{\infty} \epsilon_n \psi (z_n) \delta(z-nd),
\label{KIKS1}
\end{equation}
where $d$ is the distance between nearby scatterers, $\psi$ is the electric field of the 
TE mode, and the energy is given by $E=k^2$. We can rewrite the above equation in the discrete form for 
$\psi_n\equiv \psi(z_n = nd)$
\begin{equation}
\psi_{n+1}+\psi_{n-1} = \left[ 2 \cos(kd)-U_n\cdot kd \sin(kd) \right] \psi_n 
\label{KIKS}
\end{equation}
One can split the potential $U_n$ into a mean $\epsilon$ and a fluctuating 
term $\epsilon_n$, $U_n = \epsilon + \epsilon_n$. Eq.~(\ref{KIKS}) is then equivalent 
to the tight-binding equation (\ref{TBE}),
with energy $E\rightarrow 2 \cos k +k \epsilon \sin k$ and random potential $V_n \rightarrow k 
\epsilon_n \sin k$.

\begin{figure}[htb]
\includegraphics[width=\columnwidth,keepaspectratio,clip]{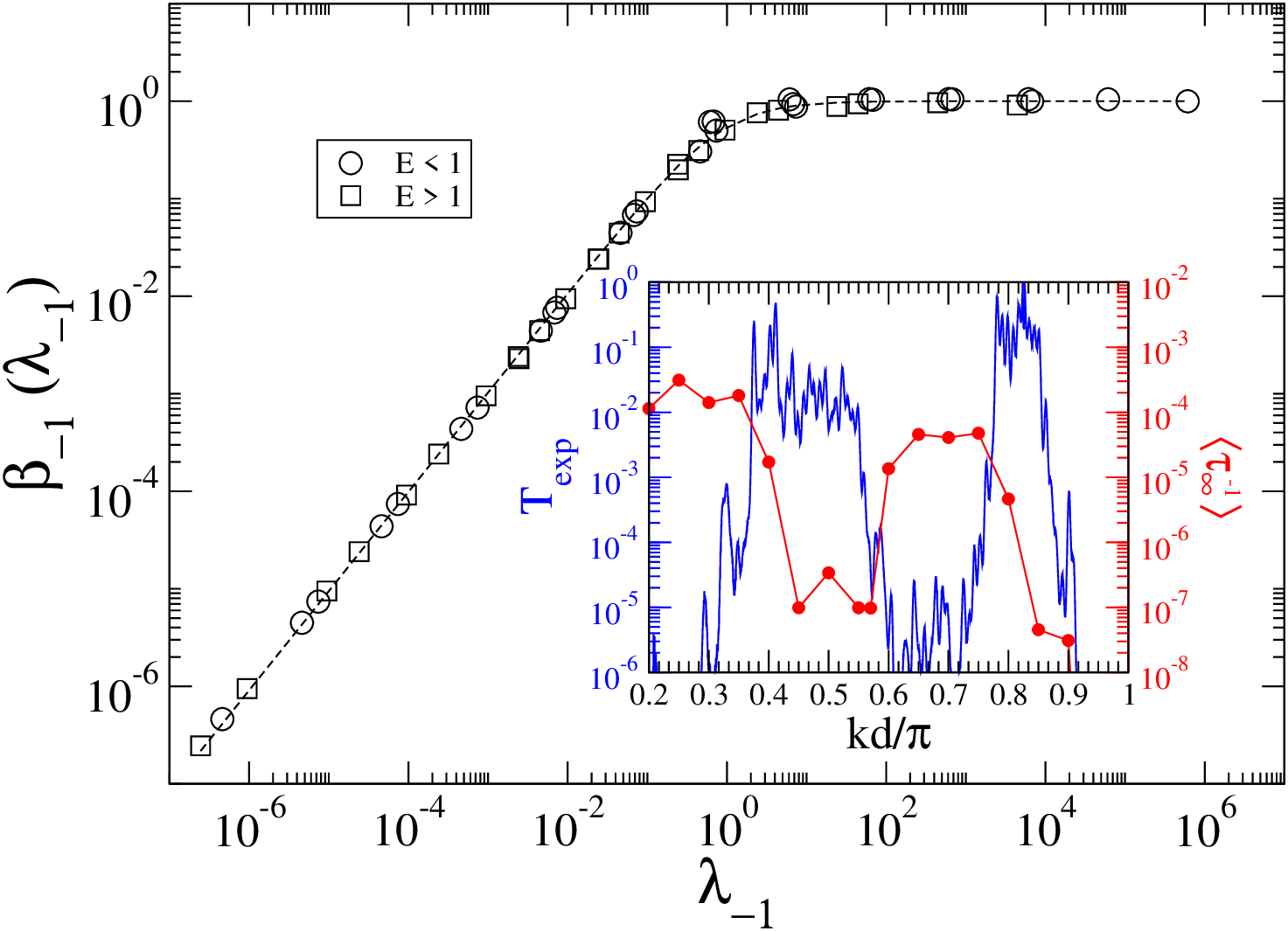}
\caption{\label{KPCorrScale} (Color online) Scaled inverse delay times for microwaves propagating in $1D$ waveguide.
The different symbols correspond to energies $|E(k = 0.5\pi)| < 1, |E(k = 0.7\pi)| > 1$ being on both sides of the critical 
wavevector $k=0.57\pi$. A nice data collapse is observed, indicating that in both cases, the statistical properties 
of delay times (and thus the structural properties of wavefunctions) are unaffected by the correlation and correspond 
to exponentially localized wavefunctions, albeit the localization length for $k = 0.5\pi$ is much larger than for 
$k = 0.7\pi$. This is reflected in the overall scaling parameter $\langle \tau_{\infty}^{-1} \rangle$.
The dashed line is the result of the best fit from Eqs. (\ref{XYvar},\ref{XYline}).
Inset: The experimental transmission coefficient showing pass and stop bands is displayed by the blue line (left axis). 
The values for $\langle \tau_{\inf}^{-1} \rangle$ are shown by the red circles (right axis).  \cite{thanks}.}
\end{figure}

By choosing the on-site potential as \cite{IK99,KIK07}
\begin{equation}
\epsilon_n = \epsilon \sum_{m=-\infty}^{\infty} \xi_m \cdot \zeta_{n+m} 
\label{AnderCorr}
\end{equation}
where $\zeta_{n+m}$ is a random variable, uniformly distributed  within the interval $(0,1]$, and 
\begin{eqnarray}
\xi_m = \left\lbrace   \begin{array}{cc}
    \sqrt{2\over \pi}(\mu_2-\mu_1)^{3/2}, & m=0 \\
    {1\over m}\sqrt{\frac{\mu_2-\mu_1}{2 \pi}}  \left[ \sin(2m \mu_2)-\sin(2m \mu_1)\right] , & m\neq 0
  \end{array} \right.  \label{xibuild}
\end{eqnarray}
with $\mu_1=0.2\pi$, $\mu_2=0.4\pi$, and $\epsilon=-0.1$ it was argued that mobility edges can 
be tailored at wavenumbers $kd/\pi = 0.38, 0.57$ and $0.76$. The experimental data \cite{KIKS00} 
(see blue line (left axis) within the inset of Fig.~\ref{KPCorrScale} \cite{thanks}) did indeed 
seem to confirm the theoretical predictions. However, various questions still remain to be 
clarified - the most prominent being the nature of the corresponding eigenstates and how they 
are structurally affected by these potential correlations.

With the help of the iteration relations, Eqs.(\ref{Taun},\ref{Phin}), we have investigated the scaling 
properties of $\langle \tau_L^{-1}\rangle$ for the correlated model, Eqs.(\ref{KIKS1},\ref{AnderCorr}). Two
energies $E$ from both sides of the mobility edge $k=0.57\pi$ have been chosen. In Fig.~\ref{KPCorrScale} 
we report our numerical data by referring to the scaling variables $\beta_{-1}$ and $\lambda_{-1}$, defined 
in Eq.~(\ref{tscale1}). The data correspond to various system sizes $L \in {10^1, 10^2,...,10^6,10^7}$
 and disordered strengths $\epsilon \in \left\lbrace 0.1, 0.5, 2.5, 5 \right\rbrace $. The remarkable 
agreement between the data from both sides of the "mobility" edge confirms again the theoretical 
prediction, Eq.(\ref{tscale1}), and indicates clearly that the corresponding eigenfunctions have 
the same {\it structural} properties, thus being unaffected by the potential correlations. Using 
the scaling properties of the Wigner delay times, we are able to conclude that $k=0.57\pi$ does 
not correspond to any mobility edge separating extended from exponentially localized eigenstates. 
Rather in both energy regimes the eigenstates are structurally the same (\textit{i.e.} exponentially 
localized), albeit the localization length is drastically different. This is reflected in the 
overall scaling factor $\langle \tau_{\infty}^{-1}\rangle$ (used to scale the data according to 
Eq. (\ref{tscale1})), illustrated by the red circles (right axis) within the inset of Fig.~\ref{KPCorrScale}. 
Note that $\tau_{\infty}\sim l_{\infty}$ (see for example \cite{OKG00}). As we can see from Fig. 
\ref{KPCorrScale}, at the pass-band region, $\langle \tau_{\infty}^{-1}\rangle$ is much smaller 
than that of the stop-band region; {\it i.e}, $l_{\infty}$ is much larger in the former case, but 
nonetheless remains finite (a "true" transition would imply that $\langle \tau_{\infty}^{-1}\rangle 
\sim L^{-1}$ and thus by increasing the system size the scaling factor had to go to zero). This
abrupt change in the magnitute of $\langle \tau_{\infty}^{-1}\rangle$ arround $k\sim 0.57\pi$ is
a fingerprint of the correlations imposed to the disordered potential. Nevertheless, after rescaling 
the data the universal scaling law Eq. (\ref{tscale1}) is again satisfied.

%\begin{figure}[htb]
%\includegraphics[width=\columnwidth,keepaspectratio,clip]{KPCorrScale2}
%\caption{\label{KPCorrScale} (Color online) Scaled inverse delay times for microwaves propagating in $1D$ waveguide.
%The different symbols correspond to energies $|E(k = 0.5\pi)| < 1, |E(k = 0.7\pi)| > 1$ being on both sides of the critical 
%wavevector $k=0.57\pi$. A nice data collapse is observed, indicating that in both cases, the statistical properties 
%of delay times (and thus the structural properties of wavefunctions) are unaffected by the correlation and correspond 
%to exponentially localized wavefunctions, albeit the localization length for $k = 0.5\pi$ is much larger than for 
%$k = 0.7\pi$. This is reflected in the overall scaling parameter $\langle \tau_{\infty}^{-1} \rangle$.
%The dashed line is the result of the best fit coming out from Eqs. (\ref{XYvar},\ref{XYline}).
%Inset: The experimental transmission coefficient showing pass and stop bands is displayed by the blue line (left axis). 
%The values for $\langle \tau_{\inf}^{-1} \rangle$ are shown by the red circles (right axis).  \cite{thanks}.}
%\end{figure}

\begin{figure}[htb]
\includegraphics[width=\columnwidth,keepaspectratio,clip]{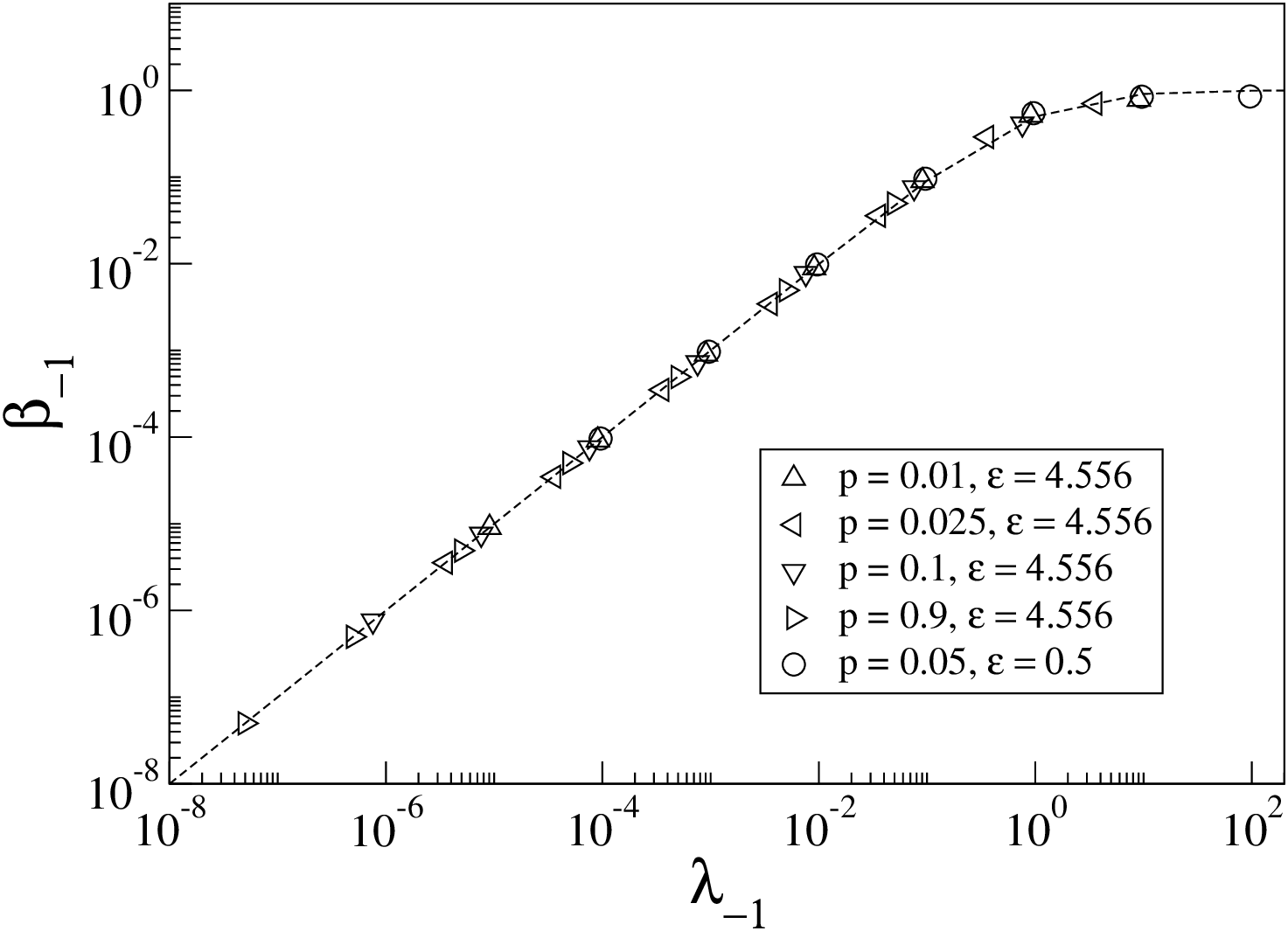}
\caption{\label{BinomScale} Scaled inverse delay time for the Disordered Optical Lattice System, 
with $\epsilon \in \left\lbrace 4.556, 0.5\right\rbrace $ and filling factor $p\in \left\lbrace 
0.01, ,0.025, 0.05, 0.1, 0.9 \right\rbrace $. The nice data collapse confirms the universality of the
scaling law, Eqs.(\ref{tscale1},\ref{tscale3}). 
The dashed line is the result of the best fit from Eqs.(\ref{XYvar},\ref{XYline}).
}
\end{figure}

%%%%%%%%%%%%%%%%%%%%%%%%%%%%%%%%%%%%%%%%%%%%%%%%%%%%%%%%
{\it Disordered Optical Lattices -} 
It was recently proposed in \cite{GC05} that we can observe Anderson localization of ultra-cold 
atoms scattered off a gas of atoms of another species or internal state, randomly trapped at 
the nodes of an optical lattice. Within this set-up, cooled vibrational ground-state atoms trapped 
at the nodes of a periodic optical lattice act as (static) delta scatterers provided that the kinetic 
energy of the incoming particles is less than the vibrational energy of the trapped scatterers
\textit{i.e.}  $\frac{\hbar^2k^2}{2m_{\rm incoming}}\ll \hbar \omega_{\rm scatterer}$. The mathematical model
that describes the motion of the incoming particle along the lattice direction is the Kronig-Penney
model, Eqs.(\ref{KIKS1},\ref{KIKS}), in this case with binary on-site potential distribution. Localization is then 
dependent on three parameters: wavevector $k$, disorder strength $\epsilon$, and the filling factor 
$p\in [0,1]$. The latter dictates a binomial distribution of the on-site potential
\begin{equation}
\epsilon_n = \left\lbrace   \begin{array}{cc}
    \epsilon, & \zeta_n < p \\
    0 , & \zeta_n \geq p
  \end{array} \right.  \label{filling}
\end{equation}
where $\zeta_n$ is a random number given by a uniform distribution and $\epsilon$ is the disorder
strength \cite{GC05}. In the numerical simulations presented in Fig.~\ref{BinomScale}, we used 
disorder strengths $\epsilon \in \left\lbrace 4.556, 0.5 \right\rbrace $ and filling factors 
$p\in \left\lbrace 0.01, 0.025, 0.05, 0.1, 0.9 \right\rbrace $. The larger disorder strength
corresponds to numerical values used in \cite{GC05}. The very nice overlap of the scaled delay 
times are once more in excellent agreement with the universality of the scaling law, 
Eqs.(\ref{tscale1}, \ref{tscale3}).

%

%%%%%%%%%%%%%%%%%%%%%%%%% Systems of Interest %%%%%%%%%%%%%%%%%%%%%%%%%%%%%%%%%%%%%%%%%%%%%

\begin{figure}[ht]
\includegraphics[width=\columnwidth,keepaspectratio,clip]{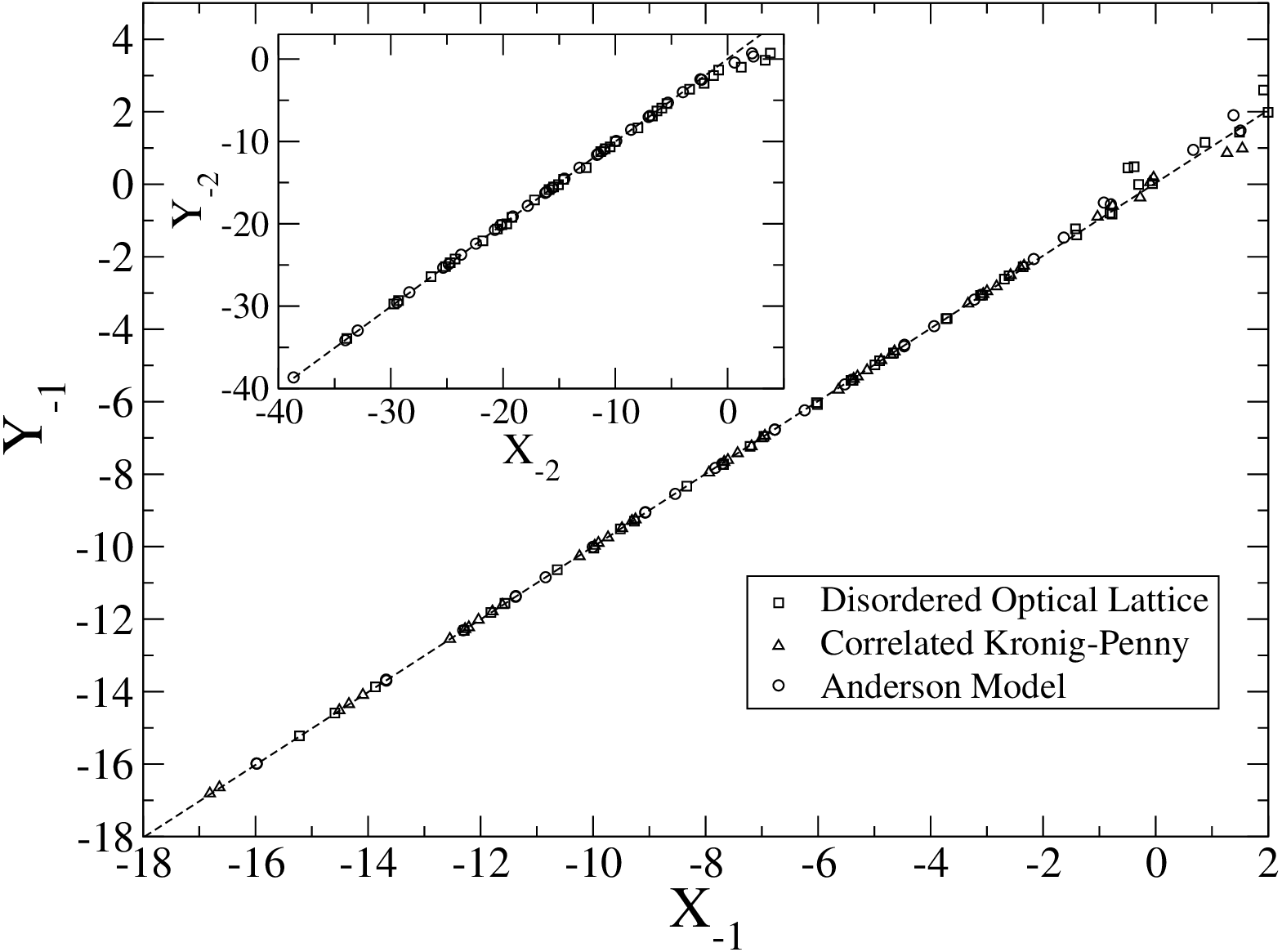}
\caption{\label{XY} Scaling of Eq.(\ref{tscale1},\ref{tscale3}) in the variables from Eq.(\ref{XYvar}).
Inset: Same as in the main figure but now for the $q=2$ case.}
\end{figure}

{\it Universal Behavior -}
It is illuminating to plot all our numerical data in the new variables 
\begin{equation}
Y_{-q} = \ln \left( \frac{\beta_{-q}}{1-\beta_{-q}} \right), \quad X_{-q} = \ln \left( \lambda_{-q} 
\right). 
\label{XYvar}
\end{equation}
In these variables the scaling for $q=1$ has an extremely simple form
\begin{equation}
Y_{-1} = a_{-1} + b_{-1} X_{-1},
\label{XYline}
\end{equation}
with $a_{-1}\approx 0$ and $b_{-1}\approx 1$. The data for the scaling in variables $Y_{-1},
X_{-1}$ are presented in Fig.~\ref{XY}. The remarkable result is that the above simple scaling 
relation holds in a very large region of the scaling parameter, $\Delta X_{-1}\approx 14$. In 
fact, Eq.(\ref{XYline}) is exact only for $q=1$, corresponding to $q'=2$ \cite{FM94,FM92}. 
However, for other values of $q$, Eq.(\ref{XYline}) is still a good approximation (see inset 
of Fig.~\ref{XY} for the case $q=2$). Placing Eq.(\ref{XYvar}) into Eq.(\ref{XYline}), we 
find that $\frac{\beta_{-1}}{1-\beta_{-1}}= \lambda_{-1}$ (see dashed lines in Figs.~\ref{AnderScale},
\ref{KPCorrScale},\ref{BinomScale}). This takes the form (\ref{tscale3}) once we substitute 
for $\beta_{-1}$ and $\lambda_{-1}$ the expressions in Eq.(\ref{tscale1}). In the inset of 
the same figure we also report the $q=2$ moment of delay times by making use of the variables 
of Eq.(\ref{XYvar}). The nice data collapse reconfirms the validity of Eq.(\ref{XYline}) 
where again $a_{-2}\approx 0$ and $b_{-2}\approx 1$ (note however that for $q=2$ small 
deviations from the straight line are evident around $X_{-2}\approx 0$)

In conclusion, we have investigated the scaling properties of inverse moments of Wigner delay 
times. We have shown that they are dictated by the scaling law, Eq.(\ref{tscale1}), which can 
be rewritten in a more familiar way, Eq.(\ref{tscale3}), resembling the scaling relation for 
the information lengths of wavefunction components. Our theoretical arguments have been tested 
in various physical models where the applicability of the non-linear $\sigma$ model is questionable, 
thus strongly supporting the relation between wavefunction moments and inverse moments of 
Wigner delay times.

TK and JDB acknowledge an Academic Excellence grant from SUN and 
thank greatly Ulrich Kuhl for inset data of Fig.\ref{KPCorrScale},
as well as insightful discussions.
JAMB thanks support from project CB-2006-01-60879, CONACyT Mexico. 
Computer time at Wesleyan University supported by the
NSF under grant number CNS-0619508.
This research was supported by a grant from the United States-Israel 
Binational Science Foundation (BSF), Jerusalem, Israel. 
%%%%%%%%%%%%%%%%%%%%%%%%%%%%%%%%%%%%%%%%%%%%%%%%%%%%%%%%%%%%%%%%%
%%%%%%%%%%%%%%%%%%%%%%%%%%%%%%%%%%%%%%%%%%%%%%%%%%%%%%%%%%%%%%%%%

%%%%%%%%%%%%%%%%%%%%%%%%%%%%%%%%%%%%%%%%%%%%%%%%%%%%%%%%%%%%%%%%%
%%%%%%%%%%%%%%%%%%%%%%%%%%%%%%%%%%%%%%%%%%%%%%%%%%%%%%%%%%%%%%%%%

\end{document}